\documentclass[10pt]{article}
\usepackage[dvips]{graphicx}

\usepackage{epsfig}
\usepackage{graphicx}
\usepackage{indentfirst}
\usepackage{amsmath}
\usepackage{amsfonts}
\usepackage{amssymb}

\begin{document}

\begin{center}
{\Large\bf   Transition between phantom and non-phantom phases with time dependent cosmological constant and Cardy-Verlinde formula\\}
\medskip

 S.J.M. Houndjo\footnote{e-mail:
sthoundjo@yahoo.fr}
\medskip

Instituto de F\'{i}sica, Universidade Federal da Bahia, 40210-340, Salvador, BA, Brazil
\medskip

\end{center}

\begin{abstract}
We investigate the transition phenomenon of the universe between a phantom and a non-phantom phases. Particular attention is devoted to the case in which the cosmological constant depends on time and is proportional to the square of the Hubble parameter. Inhomogeneous equations of state are used and the equation of motion is solved. We find that, depending on the choice of the input parameters, the universe can transit from the non-phantom to the phantom phase leading to the appearance of singularities. In particular, we find that the phantom universe ends in the singularity of type III, unlike the case without variable cosmological constant in which the phantom phase ends exclusively in the big rip (singularity of type I). The Cardy-Verlinde formula is also introduced for inhomogeneous equation of state and we find that its equivalence with the total entropy of the universe, coming from the Friedmann equations, occurs only for special choice of the input parameter $m$ at the present time.

\end{abstract}

Pacs numbers: 98.80.Es, 95.36.+X, 05.70.-a

\section{Introduction}
Accelerated expansion seems to play an important role in the dynamical history of the universe. There is a firm belief, at present time, that universe passed through inflationary phase at early times and there are growing evidences that it is accelerating at present. The study of large scale structures indicates that the universe is almost spatially flat and that dark energy accounts for about $70$ percent of the total energy content \cite{sal1,sal2,sal3,sal4}. Moreover, dark energy is believed to be the responsible of the acceleration of our expanding universe. This sort of fluid violates the strong version of the energy conditions. Besides, when the null version is also violated, the fluid is called phantom and then the universe may present future singularities at finite time \cite{stephane,shin,robert}. In the phantom phase, the energy density grows whereas it decreases in a non-phantom one. However, we know little about the nature of dark energy in general and of phantom fluid in particular, except for their negative pressure. Therefore, a large effort has been spent in recent years to explain this mystery \cite{saulo1,shinichi,salvatore,mano1,mano2,mano3}. Specially, this transition phenomenon known as quintom scenario is proposed by B. Feng and collaborators in \cite{feng}, and they found that it gives rise to the equation of state larger than $-1$ in the past and less than $-1$ today, satisfying current observations. For understanding the possible connections among the dark energy models, it is useful to study the cosmic duality. Then, it was studied the duality in two-field quintom models of dark energy and it has been found that an expanding universe dominated by quintom-A field is dual to a contracting universe with quintom-B field \cite{yifu2}. Recently, Yi-Fu. Cai and collaborators wrote a paper which introduced the experimental developments on finding the transition between quintessence and phantom phases and various theoretical realizations of such a scenario, see \cite{yifu}. On the other hand, the knowledge of some properties of the universe in the phantom phase, such as the singularities, motivated various investigations with the aim of dealing with them. A. B. Batista and collaborators \cite{brasil} investigated the effects of particle production when a massless minimally coupled scalar field is present in spacetimes where $\omega$ is a constant. To do so they used a state for which Bunch and Davies \cite{bunch} had previously computed the stress-energy tensor. They found that the energy density of the created particles never dominates over the phantom energy density. In the same way, quantum effects near the big rip is studied in \cite{flavio} where they used the n-wave regularization for calculating the energy density of particle creation and found that, in this case, it tends to infinity when the big rip is approached and becomes the dominant component of the universe. This means that the big rip can be avoided by a scalar massless field. Pavlov \cite{pavlov} computed both the number density of created particles and the stress-energy tensor for a conformally coupled massive scalar field for the case in which $\omega = -5/3$. It was found that quantum effects are not important for masses much smaller than the Planck mass and times which are early enough that the time until the Big Rip occurs is greater than the Planck time. J. D. Bates and P. R. Anderson \cite{anderson} used a background field approach in which the energy densities of the quantized fields are computed in the background spacetime which contains the Big Rip singularity. They found that for fields in realistic states for which the energy density of the quantized fields is small compared to that of the phantom energy density at early times, and for spacetimes with realistic values of $\omega$, there is no evidence that quantum effects become large enough to significantly affect the expansion of the spacetime until the spacetime curvature is of the order of the Planck scale or larger, at which point the semi classical approximation breaks down. Also in order to deal with singularity problem, it has been considered by Yi-Fu Cai and collaborators \cite{yifu3} the cosmology of the Higgs sector of the Lee-Wick Standard Model, an alternative to supersymmetry to solving the hierarchy problem. They found that homogeneous and isotropic solutions are non-singular and then, the Lee-Wick model can provide a possible solution of the cosmological singularity problem.
 \par
One of the phenomenological ways to explain the dark energy problem is assuming a variable cosmological constant. The cosmological constant $\Lambda$ is pretty compatible with observation and effort is currently devoted to the investigation of the theoretical foundations of a variable cosmological constant and its model properties \cite{strominger,dymnikova1,dymnikova2,khlopov}. In this respect, various ansatz have been used. For example in \cite{saulo2}, S. Carneiro and collaborators considered a cosmological constant $\Lambda \propto H$, and studied a possible way to distinguish the validity of this scenario from the standard one. Note that a model $\Lambda \propto a^{-2}$ has earlier been proposed in \cite{ozer}, requiring that the cosmic density $\rho$ equal to the Einstein-de Sitter critical density $\rho_c$, leading to a close universe without singularity, horizon, entropy and monopole problems. Furthermore, a large number of phenomenological $\Lambda$ models have been constructed in order to describe the dynamics of the universe \cite{overduin}. The case which attracts our attention in this paper is $\Lambda\propto H^{2}$ and has been studied in several other works for other purposes, but always with the aim of clarifying some grey areas in cosmology \cite{freese,carvalho,lima1,lima2}.  In this paper we propose to use $\Lambda\propto H^2$, for analysing  the evolution of the universe, being in the phantom or non-phantom phase. The phase transition of the universe will be studied considering a model in which dark energy is described by some rather complicated ideal fluid  with an unusual equation of state (EoS) which will be chosen to be an inhomogeneous one. It is important to emphasize that this inhomogeneous EoS corresponds to a pure dark energy models. This kind of models may reproduce late-time acceleration, but it is not easy to construct a model that keeps untouched the radiation and matter dominated epochs. However, it is easy to introduce into our considerations ordinary matter and radiation but in that case they only appear suddenly at some point.

 Note that this kind of study has been also considered in \cite{brevik2} where the cosmological constant is a linear function  of time. Here, we use the ansatz $\Lambda \propto H^2$ with the inhomogeneous EoS. We find that, depending one the choice of the input parameters of the EoS considered, the universe may transit from a non-phantom to a phantom phase leading to finite time singularities. Another interesting point of our result is that, in contrast to the model without variable cosmological constant in which the phantom universe ends with the singularity of type I (big rip), the phantom universe in this case ends with the singularity of type III.
\par
Another point we address here is the Cardy-Verlinde (CV) formula coming from inhomogeneous EoS. Verlinde \cite{verlinde} made an interesting proposal that Cardy formula \cite{cardy} in two-dimensional conformal field theory can be generalized to arbitrary spacetime dimensions. Verlinde further proposed that a closed universe has subextensive (Casimir) contribution to its energy and entropy with the Casimir energy conjectured to be bounded from above
by the Bekenstein-Hawking energy and as consequence, one obtains a very deep relation between gravity and thermodynamics \cite{youm}. Within the context of the radiation dominated universe, such bound on the Casimir energy is shown to lead to the Hubble and the Bekenstein entropy bounds respectively for the strongly and the weakly self-gravitating universes. The generalized entropy formula, called the CV formula, is further shown to coincide with the total entropy of the universe coming from the Friedmann equations. These results were later generalized \cite{group1,group2,group3,group4,group5,group6,group7,group8}. Our goal here is to analyse the equivalence between the CV formula and the total entropy coming from Friedmann equations assuming that the universe is conformally invariant. With this, we find that for the inhomogeneous EoS, the generalized entropy of the universe reduces to the CV formula with a special choice of the input parameter $m$ and this does not correspond to a radiative universe as in the case of homogeneous EoS. Note also that this equivalence occurs exclusively at the presente time .\par
The paper presents two sections, the first showing the inhomogeneous EoS with which the solution for the Hubble parameter is found and some discussions on the input constants are put forward, allowing to a whole analysis of the transition phenomenon. The second section shows a brief concept on the CV formula with homogeneous EoS and latter, a complete analysis of the equivalence between the CV formula  and the Friedmann equations with inhomogeneous EoS. Finally, we present our conclusions and perspectives.

\section{Solving the inhomogeneous equation of state}
Les us consider the universe driven by an ideal fluid (dark energy) with the inhomogeneous equation of state \cite{brevik1}
\begin{equation}\label{e1}
p=\omega(t)\rho+\Lambda(t)\quad,
\end{equation}
where $\omega(t)$ and $\Lambda(t)$ depend on the time and $\rho$ and $p$ are respectively the energy density and the pressure of the fluid.  This equation, for the case $\Lambda(t)=0$ and $\omega(t)$ as affine function of time, has been studied in \cite{brevik1,nojiri1}. Moreover, the case $\Lambda(t)\neq 0$ as a affine function of time has been examined in \cite{brevik2}.\par
The equation of energy conservation and the Friedmann equations are respectively
\begin{eqnarray}
\dot{\rho}+3H\left(\rho+p\right)=0\label{e2}\quad,\\
\frac{3}{\kappa^2}H^2=\rho\label{e3}\quad,\\
\frac{1}{\kappa^2}\left(2\dot H+3H^2\right)=-p\label{pressure}\quad,
\end{eqnarray}
where $\kappa^2=8\pi G$, with $G$ the gravitational constant and $H=\frac{\dot a}{a}$, the Hubble parameter where $a(t)$ is the scale factor. \par
Using (\ref{e1}) and (\ref{e3}), equation (\ref{e2})  is rewritten as 
\begin{eqnarray}
\dot{\rho}+\sqrt{3}\kappa\left[1+\omega(t)\right]\rho^{3/2}+\sqrt{3}\kappa\rho^{1/2}\Lambda(t)=0\label{e4}\quad.
\end{eqnarray}
From now on, we suppose that $\omega(t)$ depends linearly on time and the cosmological constant $\Lambda(t)$ is proportional to the square of the Hubble parameter, that is
\begin{eqnarray}
\omega(t)&=& \alpha t+\beta \label{e5}\quad,\\
\Lambda(t)&=&\gamma H^2(t)\nonumber\\
&=&\frac{\gamma\kappa^2}{3}\rho(t)\label{e6}\quad.
\end{eqnarray}
Taking into account (\ref{e5}) and (\ref{e6}), equation (\ref{e4}) becomes
\begin{eqnarray}\label{e7}
\dot{\rho}+\left(At+B\right)\rho^{3/2}=0\quad,\quad A=\alpha\kappa\sqrt{3}\,\,\,,\quad B=\kappa\sqrt{3}\left(1+\beta+\frac{\gamma\kappa^2}{3}\right)\,\,\,.
\end{eqnarray}
The solution of this equation is
\begin{equation}\label{e8}
\rho(t)=\frac{16}{\left(At^2+2Bt-2C\right)^2}\,\,\,,
\end{equation}
and the Hubble parameter and its rate behave as
\begin{equation}\label{e9}
H(t)=-\frac{4\kappa}{\sqrt{3}\left(At^2+2Bt-2C\right)}\,\,\,,
\end{equation}
\begin{eqnarray}\label{e10}
\dot{H}(t)=\frac{8\kappa\left(At+B\right)}{\sqrt{3}\left(At^2+2Bt-2C\right)^2}\,\,,
\end{eqnarray}
where $C$ is an integration constant.\par
As we are dealing  with an expanding universe, we need an increasing scale factor, that is $\dot{a}>0$. We know that $\dot{a}= Ha$, then for an expanding universe the Hubble parameter has also to be positive. Solving $\dot{a}= Ha$, one obtains 
\begin{eqnarray}
a(t)&=&\exp{\left(\int H(t)dt\right)}\nonumber\\
&=&\exp{\left(\frac{4\kappa}{\sqrt{3}}\frac{g(t)}{\sqrt{B^2+2AC}}\right)}\,\,,\quad g(t)=\arctan{\left(\frac{At+B}{\sqrt{B^2+2AC}}\right)}\label{e11}\,\,.
\end{eqnarray}
The positivity of $\dot{a}$ depends on the sign of $H(t)$. Note that the expression $At^2+2Bt-2C$ vanishes for $t_{1,2}=-\left(B\pm\sqrt{B^2+2AC}\right)/A$.  The universe expands when the Hubble parameter is positive and one has a phantom fluid when the weak  version of the energy conditions is violated, that is when $\rho+p<0$. Combining equations (\ref{e3}) and (\ref{pressure}), one obtains
\begin{eqnarray}
\rho+p=-\frac{2}{\kappa^2}\dot{H}\label{e12}\,\,\,,
\end{eqnarray} 
and it turns out that the phantom  phase (respectively the non-phantom one) is obtained when $\dot{H}>0$ ($\dot{H}<0$). Two situations are important for a whole analysis: when the constant $A$ is positive or negative.\par
$\bullet$ {\bf Analysis for the case $A>0\, (\alpha>0)$}\par
 In this case, a simple study of the Hubble parameter sign shows that one has an expanding universe for $t_1<t<t_2$  and a contracting one when $t<t_1$ and $t>t_2$. On the other hand, it is easy to see that $\dot{H}$ vanishes for $t_3=-B/A$. Then, the universe is in the phantom phase ($\dot{H}>0$) when $t>t_3$  and the energy density grows; for the non-phantom phase ($\dot{H}<0$) $t<t_3$, the energy density decreases. Consequently, the accelerated expanding universe begins with a non-phantom phase and enters in the phantom one at the transition time $t_{tr}=t_3$, the time at which the Hubble parameter and the energy are 
\begin{eqnarray}
H_{tr}= \frac{4\kappa A}{\sqrt{3}\left(B^2+2AC\right)},\quad \rho_{tr}=\frac{16A^2}{\left(B^2+2AC\right)^2}\,\,\,.
\end{eqnarray}
In the non-phantom case, the energy density decreases and tends to $\rho_{tr}$  as the time goes to  $t_{tr}$. 
The simultaneous divergence of $\rho(t)$ and $H(t)$ appears at $t_1$ and $t_2$. However, $a(t_2)=\exp{\left(   \kappa\pi/\sqrt{3(B^2+2AC)} \right)}$, which is finite. In fact, the phantom universe ends with a future singularity, in this case, the singularity is of type III (for a classification of future singularities see \cite{stephane}) since the energy density and the pressure at this time are divergent. The graph of $H(t)$ is shown in the left panel of Fig. $1$.\par
$\bullet$ {\bf Analysis for the case $A<0 \,(\alpha<0)$}\par
For $t<t_1$ and $t>t_2$, the universe expands whereas it contracts for $t_1<t<t_2$. The first derivative of the Hubble parameter is positive (respectively negative) for $t<t_3$ $(t>t_3)$. The expanding universe begins with a phantom phase which ends at $t_1$ and enters in a non-phantom phase at $t_2$. Here the transition is not instantaneous, it is the contracting phase of the universe. At $t_1$, the energy density $\rho(t_1)$ and the pressure $p(t_1)$ diverge. However, at $t_1$, the scale factor is finite, $a(t_1)=\exp{\left(  - \kappa\pi/\sqrt{3(B^2+2AC)} \right)}$. Then, the phantom phase ends with the singularity of type III. In the non-phantom phase, the energy density decreases and goes to zero as $t\longrightarrow \infty$. The graph of the Hubble parameter, $H(t)$,  versus time $t$ is shown in the right panel of Fig. $1$. \par

 \par
\section{ CV formula from inhomogeneous EoS fluid}
This section is devoted to the application of CV formula to ideal fluids. In a first step, let us tell briefly introduce CV formula for homogeneous EoS. We consider a (n + 1)-dimensional space time described by the FRW metric, written in comoving coordinates as
\begin{eqnarray}\label{cv1}
ds^2=dt^2-\frac{a^2(t)dr^2}{1-kr^2}-r^2d\Omega^{2}_{n-1}\,\,,
\end{eqnarray} 
where $k = -1, 0, +1 $ for an open, flat, or closed spatial Universe respectively, and $d\Omega^{2}_{n-1}$ is the metric of an $n-1$
sphere. Then, by inserting the metric (\ref{cv1}) in the Einstein equations the Friedmann equations are derived,
\begin{eqnarray}\label{cv2}
H^2=\frac{16\pi G}{n(n-1)}\rho-\frac{k}{a^2}\,\,\,, \quad\quad \dot{H}= -\frac{}{}\left(\rho+p\right)+\frac{k}{a^2}\,\,.
\end{eqnarray}
The total energy $E$ of the universe is $E=\rho V$, with $V$ its total volume. Since the Casimir energy may be include in the total energy, we consider a closed universe, $k=1$. For the homogeneous EoS, $p=\omega \rho$, with $\omega$ a constant. Then the conservation law for energy has the form
\begin{eqnarray}\label{cv3}
\dot{\rho}+nH\left(1+\omega\right)\rho=0\quad,
\end{eqnarray}
which reduces to (\ref{e2}) for $n=3$ and whose solution depends on the scale factor as
\begin{eqnarray}\label{cv4}
\rho \propto a^{-n(1+\omega)}\quad.
\end{eqnarray}
The total energy of the universe can be written as the sum of an extensive part $E_E$ and a subextensive part $E_C$, called the Casimir energy, and it takes the form:
\begin{eqnarray}\label{cv5}
E(S,V)= E_E(S,V)+\frac{1}{2}E_C(S,V)\,\,\,.
\end{eqnarray}
Under the transformations $S\rightarrow \xi S$ and $V\rightarrow \xi V$ with a constant $ \xi $, the extensive and
the subextensive parts of the total energy respectively scale as \cite{sergei1,cai} 
\begin{eqnarray}\label{cv6}
E_E(\xi S,\xi V)= \xi E_E(S,V),\quad E_C(\xi S,\xi V)= \xi^{1-\frac{2}{n}}E_C(S,V)\quad,
\end{eqnarray}
and therefore, we have for the total energy
\begin{eqnarray}\label{cv7}
E(\xi S, \xi V)= \xi E_E(S,V)+\frac{1}{2}\, \xi^{1-\frac{2}{n}}E_C(S,V).
\end{eqnarray}
Taking the derivative of (\ref{cv7}) with respect to $\xi$ and letting $\xi=1$, one obtains
\begin{eqnarray}\label{cv8} 
S\left(\frac{\partial E}{\partial S}\right)_V+V\left( \frac{\partial E}{\partial V}\right)_S = E_E+\left(\frac{1}{2}-\frac{1}{n}\right)E_C\,\,.
\end{eqnarray}
Assuming that the universe satisfies the first law of thermodynamics $dE = TdS-pdV$, we have the thermodynamics relations $\left(\frac{\partial E}{\partial V}\right)_S= - p$ and $\left(\frac{\partial E}{\partial S}\right)_V = T$. Using these thermodynamics relations and Eq. (\ref{cv5}), one can put Eq. (\ref{cv8}) into the following form for the Casimir energy, as the violation of the Euler identity
\begin{eqnarray}\label{cv9} 
E_C= n \left( E+pV- TS\right)\,\,.
\end{eqnarray}
Since the total energy behaves as $E \sim a^{-n\omega}$ and by Eq. (\ref{cv5}), the Casimir energy also goes as $E_C \sim a^{-n\omega}$. The FRW Universe expands adiabatically ($dS = 0$) so the products $E_C a^{n\omega}$ and $E_E a^{n\omega}$ should be independent of the
volume V , and be just a function of the entropy. Then, by the rescaling properties (\ref{cv6}), the extensive and subextensive
parts of the total energy can be written as functions of the entropy only \cite{youm},
\begin{eqnarray}\label{cv10}
E_E=\frac{\mu}{4\pi a^{n\omega}}S^{\omega+1}\,\,,\quad\quad E_C=\frac{\nu}{2\pi a^{n\omega}}S^{\omega+1-2/n}\,,
\end{eqnarray}
where $\mu$ and $\nu$ are undetermined constants and $4\pi$ and $2\pi$ are used for convenience. From these expressions for $E_E$ and $E_C$, one obtains the following expression for the entropy of the universe:
\begin{eqnarray}\label{cv11}
S= \left(\frac{2\pi a^{n\omega}}{\sqrt{\mu\nu}}\sqrt{E_C(2E-E_C)}\right)^{\frac{n}{n(\omega+1)-1}}\,.
\end{eqnarray}
This result, obtained in \cite{youm}, reduces to the CV formula when the universe is radiation dominated, $\omega = 1/n$, that is
\begin{eqnarray}\label{cv12}
S= \frac{2\pi a}{\sqrt{\mu\nu}}\sqrt{E_C(2E-E_C)}\,.
\end{eqnarray} 
Let us now look to the case of the inhomogeneous EoS that we used in the precedent section and analyse the relationship between the CV formula and the entropy of the universe. As has been done in \cite{brevik12}, we assume an EoS expressed as a function of the scale factor and described by
\begin{eqnarray}\label{cv13}
p=\omega(a)\rho+j(a)\,\,.
\end{eqnarray}
Introducing (\ref{cv13}) in the energy conservation equation (\ref{cv3}), one obtains
\begin{eqnarray}\label{cv14}
\rho^{\,\prime}(a)+\frac{n(1+\omega(a))}{a}\rho(a)= - n\frac{j(a)}{a}\,\,,
\end{eqnarray}
where the prime denotes the derivative with respect to the scale factor and we took $t=t(a)$. The general solution of (\ref{cv14}) is
\begin{eqnarray}\label{cv15}
\rho(a)= e^{-F(a)}\left( Q-n\int e^{F(a)}\frac{j(a)}{a}da\right)\,\,, \mbox{with}\quad F(a)= n\int^{a}\frac{1+\omega(a^{\prime})}{a^{\prime}}da^{\prime}\,\,,
\end{eqnarray}
where $Q$ is an integration constant. In this analysis, making use of (\ref{e6}) and taking into account the derivative with respect to the scale factor, (\ref{e1}) can be written as
\begin{eqnarray}\label{cv16}
\rho^{\prime}(a)+\frac{n(1+\bar{\omega}(a))}{a}\rho(a)=0\,\,,\quad\quad \bar{\omega}(a)= \frac{\gamma\kappa^2}{3}+\omega(a)\,\,.
\end{eqnarray}
Identifying (\ref{cv16}) with (\ref{cv14}), one gets 
\begin{eqnarray}\label{cv17}
j(a)=0\,,\quad  F(a)= n\int^{\,a}\frac{1+\bar{\omega}({a^{\prime}})}{a^{\prime}}da^{\prime}\,\,,
\end{eqnarray}
from which one obtains the energy density as
\begin{eqnarray}\label{cv18}
\rho(a)= Q e^{-F(a)}\,\,.
\end{eqnarray}
On the other hand, using (\ref{e11}), one can write
\begin{eqnarray}\label{cv19}
\bar{\omega}(a)=\frac{\sqrt{B^2+2AC}}{\kappa\sqrt{3}}\tan{\left[ \ln{\left(a^{\sqrt{3(B^2+2AC)}/(4\kappa)}\right)} \right]}\,\,.
\end{eqnarray}
Making use of (\ref{cv16}), (\ref{cv17}), (\ref{cv19}) and (\ref{cv18}), the energy density is written as 
\begin{eqnarray}\label{cv20}
\rho(a)=Q\,\left[ \cos{\left(\frac{\sqrt{3(B^2+2AC)}}{4\kappa} \ln{(a)} \right)}\right]^{\frac{4n}{3}}\,\,.
\end{eqnarray}
Note here that only for some special conditions of the functions $\omega(a)$ and $j(a)$ CV formula (\ref{cv12}) can be recovered . Let us assume that the present time is $t_0=0$ and then analyse the equivalence between CV formula and the total entropy of the universe at this moment. Note that as $t\rightarrow t_0$, $\bar{\omega}(a)\rightarrow m-1$, with $m=1+\beta+\frac{\gamma\kappa^2}{3}$. Then, $F(a)\rightarrow nm\ln{(a)}$ and $\rho\propto a^{-nm}$. Hence, the total energy in the volume $V=a^n$ behaves as $E=\rho V\propto a^{n(1-m)}$, which is the same behaviour for the  extensive and subextensive energy through (\ref{cv9}) and (\ref{cv5}). If we assume the conformal invariance, the products $E_E a^{n(m-1)}$ and $E_C a^{n(m-1)}$ do not depend on the volume and are only functions of entropy. Then, we have for the extensive and subextensive energy,
\begin{equation} \label{cv21}
E_E= \frac{\mu}{4\pi a^{n(m-1)}}S^{m}\,\,,  \quad E_C= \frac{\nu}{2\pi a^{n(m-1)}}S^{m-\frac{2}{n}}\,\,,
\end{equation}
from which we determine the entropy as
\begin{eqnarray}\label{cv22}
S=\left[\frac{2\pi n a^{n(m-1)}}{\sqrt{\mu\nu}}\sqrt{E_C(2E-E_C)}\right]^{\frac{n}{nm-1}}\,\,.
\end{eqnarray}
Then, for $m=\frac{n+1}{n}$, the CV formula is recovered. However, for any $m\neq \frac{n+1}{n}$ CV formula can not be reproduced. Note in this case that the universe does not correspond to a radiative one.

\section{Conclusion}
We studied the transition of the universe between a phantom and a non-phantom phases. Note that in the non-phantom phase, the energy density decreases while in the phantom one, it grows leading to singularities. We focused our attention on the variable cosmological constant which has been introduced in the EoS, which becomes inhomogeneous. Then, we solved the equation of motion which led to the explicit expression of the energy density and consequently to that of the Hubble parameter. The first derivative of the Hubble parameter played a crucial role in this analysis since it allowed us to know which time interval corresponds to the phantom or non-phantom universe. The input parameter $A$ also appeared to be an important one.\par 
In the first part, with the EoS considered and the ansatz $\Lambda\propto H^2(t)$, two important cases have been found. For $A>0$, we saw that the universe evolves from a non-phantom  phase, $t_1<t<t_{tr}$, to a phantom one $t_{tr}<t<t_2$. In the non-phantom phase, the energy density descreses and goes to the energy density at the transition time while in the phantom phase, the energy density and the pressure grow and go to infinity at the finite time $t_2$. At the same time the scale factor remains finite and we conclude that the universe ends with the singularity of type III. For $A<0$, we saw that the universe begins with a phantom phase which ends with the singularity of type III at $t_1$, enters in the non-phantom at $t_2$ where the energy density descreses end goes to zero as $t\longrightarrow\infty$. Here, the transition from the phantom phase to the non-phantom one is not instantaneous; it is a time interval corresponding to the contracting phase of the universe.\par
However, in the case in which the phantom phase precedes the contracting phase of the universe, it would be interesting to study the possible avoidance of this type of singularity introducing either the viscosity term in the cosmic fluid or taking into account quantum effects. On the other hand, the same analysis can be done with the ansatz that the variable cosmological constant is proportional to the Hubble parameter, $\Lambda(t)\propto H(t)$. We will address these considerations in a future work. 
\par

We also analyse the equivalence between the CV formula and the Friedmann equations with inhomogeneous EoS. Note that this has been done in several works with the homogeneous EoS and the Friedmann equations coincide with CV formula only in radiative universe. In this work, we use the inhomogeneous EoS including the variable cosmological constant proportional to the square of the Hubble parameter. We find that the equivalence between the Friedmann equations occurs only at the present time in a special case which is not the radiative universe as for the homogeneous EoS. \par   

\vspace{1cm}

{\bf Acknowledgement:} The author thanks professors S. D. Odintsov, S. Carneiro and O. Piattella for criticism and comments, and CNPq (Brazil) for partial
financial support.

\begin{figure}[htt!!!]
\begin{minipage}[t]{0.55\linewidth}
\includegraphics[width=\linewidth]{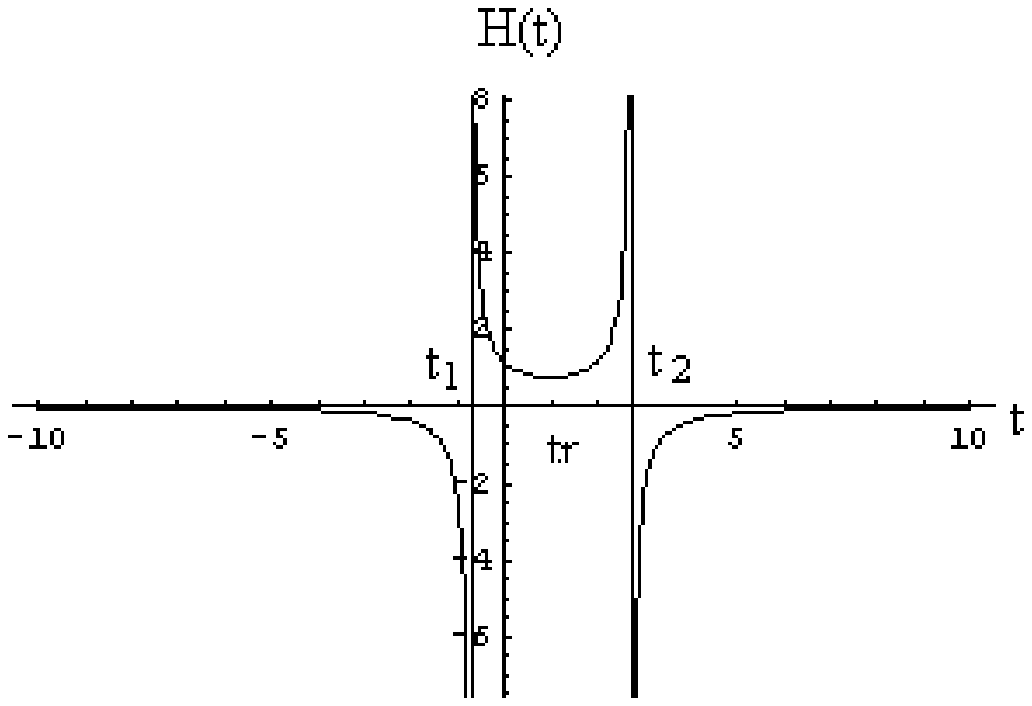}
\end{minipage} \hfill 
\begin{minipage}[t]{0.55\linewidth}
\includegraphics[width=\linewidth]{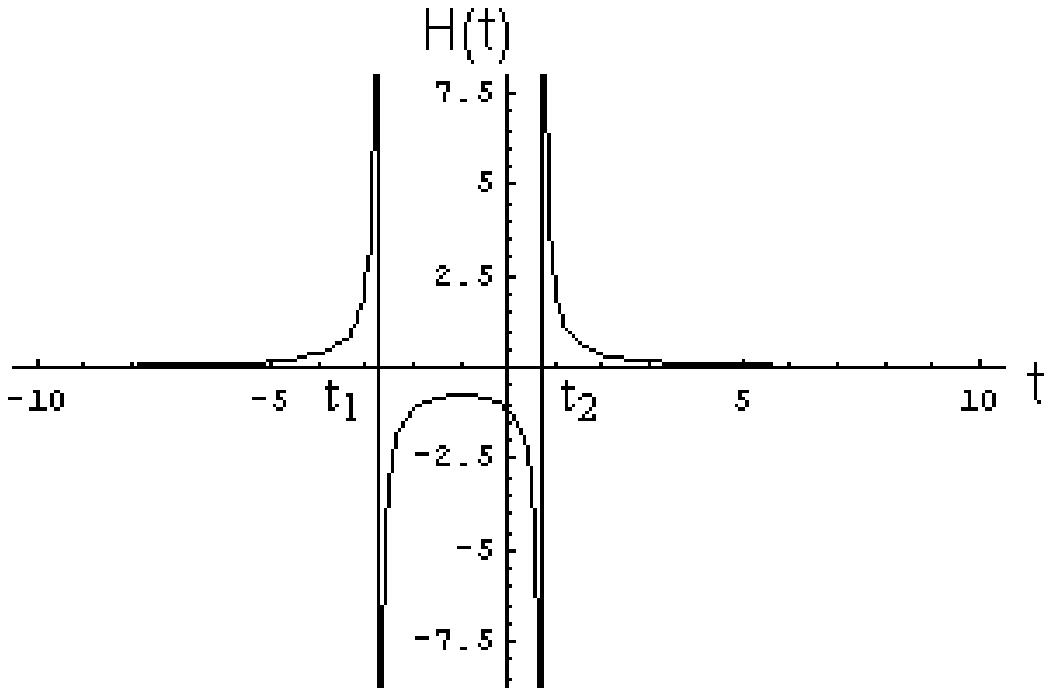}
\end{minipage} \hfill 
\caption{{\protect\footnotesize The Hubble parameter as function of the cosmic time with $B=1$, from the left to right, $A=C=1$ and $A=C=-1$ respectively.}}
\label{}
\end{figure}

\end{document}